# Characterisation of hydraulic fractures in limestones using X-ray microtomography

**François Renard** *,** — **Dominique Bernard** *** — **Jacques Desrues** **** — **Erwan Plougonven** *** — **Audrey Ougier-Simonin** *

* LGIT-CNRS-OSUG, University of Grenoble, BP 53, F-38041 Grenoble, France
francois.renard@ujf-grenoble.fr

** Physics of Geological Processes, University of Oslo, Norway

*** Institut de Chimie de la Matière Condensée de Bordeaux, ICMCB-CNRS, 87 avenue du Dr. A. Schweitzer, F-33608 Pessac, France
bernard@icmcb-bordeaux.cnrs.fr

**** Laboratoire 3S, INPG, BP 53, F-38041 Grenoble cedex 9, France
Jacques.Desrues@hmg.inpg.fr

ABSTRACT: *Hydraulic tension fractures were produced in porous limestones using a specially designed hydraulic cell. The 3D geometry of the samples was imaged using X-ray computed microtomography before and after fracturation. Using these data, it was possible to estimate the permeability tensor of the core samples, extract the path of the rupture and compare it to the heterogeneities initially present in the rock.*

RÉSUMÉ: *Des échantillons de calcaires poreux ont été fracturés dans une cellule hydraulique spécialement construite à cet effet. La géométrie 3D des échantillons a été mesurée par microtomographie synchrotron avant et après fracturation. Les données obtenues ont permis de 1) calculer le tenseur de perméabilité; 2) caractériser la rugosité des fractures produites en fonction de la distribution des hétérogénéités initialement présentes dans la roche.*

MOTS-CLÉS: *microtomographie, fracturation hydraulique, tenseur de perméabilité*

**1. Introduction**

The growth of a fracture in rocks under the driving effect of a propagating fluid arises in various geological systems. Natural fractures involve the propagation of intrusive magmatic dykes and sills in the Earth's lithosphere, and sometimes the release of magma at the Earth's surface in volcanoes. Hydraulic fracturing is used in the oil industry to stimulate oil and gas production in low permeability underground reservoirs. Other industrial applications involve the extraction of heat in geothermal reservoirs or induced caving in the mining industry.

Usually, models of hydraulic fractures consider a 2D planar fracture plane inside a homogeneous elastic medium (Bunger and Detournay, 2005; Zhang *et al.,* 2005, Garagash, 2006). However, rocks are heterogeneous at all scales. It can be shown that among other properties, the cohesion heterogeneity has a crucial effect on the fracturing pattern. Discrete simulations using elastic spring networks with disorder demonstrate that hydraulic fracture can even acquire fractal geometries, with strong spatial and temporal correlations as the fracture propagates into successive bursts clustered in time (Tzshichholz and Herrmann, 1995).

In field hydraulic experiments, two types of measurements can be performed. On one hand the pressure fluctuations can be monitored at the injection pump, and on the other hand acoustic emission signals can be recorded at the surface. Depending on the far-field stress, the injection rate and the rock permeability, either shear fractures or tension fractures are the dominant mechanism for hydraulic fracture propagation (Lockner and Byerlee, 1977). As hydraulic fractures migrate parallel to the main compressive stress, they represent a measure of the local main stress direction in the Earth's crust (Talebi and Cornet, 1987).

What is less known is how such fractures propagate in a porous medium and how the fracture tip interacts with pores and grains at small scale. Moreover, heterogeneities in the fluid flow related to local variations of the permeability can modify the stress field at the fracture tip during its propagation. Here we use a small hydraulic fracture cell to generate fractures into porous limestones. We image the core samples, using high resolution X-ray tomography, before and after fracturation. Using these 3D data, we can 1) calculate the permeability tensor of the core samples; 2) isolate the fracture path and characterise its roughness property, based on the initial heterogeneous distribution of pores in the rock. In the following sections we describe the experimental procedure and present some results on the calculation of the permeability and the roughness of hydraulic fractures.

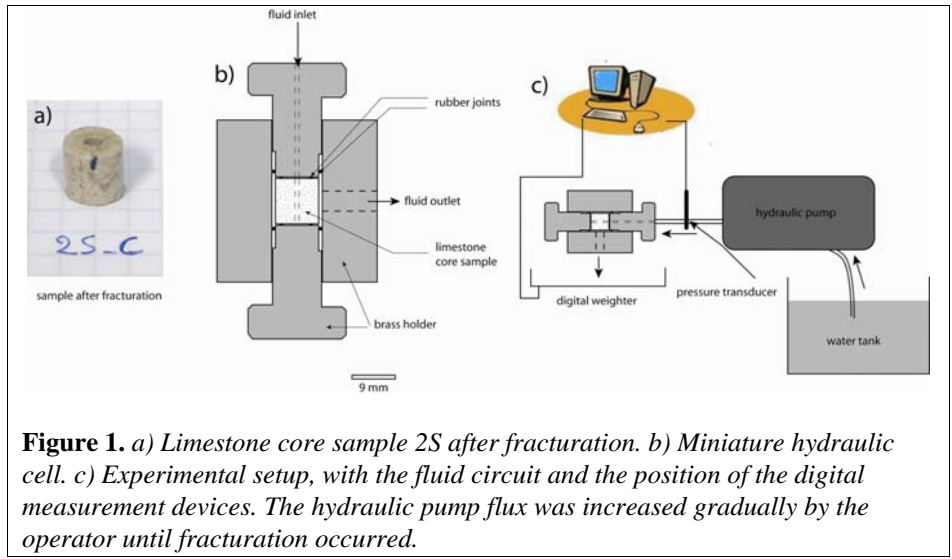

**Figure 1.** *a) Limestone core sample 2S after fracturation. b) Miniature hydraulic cell. c) Experimental setup, with the fluid circuit and the position of the digital measurement devices. The hydraulic pump flux was increased gradually by the operator until fracturation occurred.*

## 2. Experimental hydrofractures in limestones

### 2.1. Sample preparation and experimental set-up

A hydraulic cell was especially designed to allow the measurement of fracture onset in limestone core samples (Figure 1). Four samples (Table 1) were drilled from blocks of porous limestones extracted from two different quarries: 1) The Estaillades limestone contains debris of shells cemented by calcite. This is a 95% calcite rock, with 30% porosity, and permeability between 0.2 and 0.4 Darcy. Its mechanical resistance to compression is 13 to 18 MPa. 2) The Anstrude rock is an oolitic limestone with 20% porosity, a low permeability of 2 to 4 mD, and a uniaxial compression strength between 30 and 40 MPa.

In each core samples a hole was carefully drilled along the cylinder axis. The resulting sample is a hollow cylinder, with an external diameter of 9 mm fixed by the range of the CCD camera used for microtomography, and a central 2 mm diameter hole. The advantage of this geometry is that, as water is injected through the central hole, the pressure gradient had a cylindrical symmetry in the sample.

Each sample was placed between two brass holders and a rubber joint ensured a flat contact between the sample and the holder. The sample was free to deform radially. This setup allowed a homogeneous contact of the sample with the holder, to avoid fluid leakage out of the central hole (Figure 1).

*2.2. Synchrotron X-ray tomography data*

The 3D microtomographic data were acquired from beamline ID19 at the European Synchrotron Radiation Facility under energy of 25 keV and at a pixel size of 5.1 micrometers. An X-ray beam was sent onto the sample, which was rotated over a range of 180° in 1000 steps around a vertical axis, and a series of radiographs was taken using a low-noise 2048x2048 FRELON CCD camera (Fast REadout LOw Noise). Based on the sets of projections, a reconstruction algorithm was used to obtain a tomogram representing the 3D microstructure of the sample. Well-contrasted images were obtained, resulting from the difference in X-ray absorption between calcite grains and empty pores, before and after fracturation (Figure 2b, c).

| Sample | Anstrude 4M | Estaillades 1S | Estaillades 2S | Estaillades 3M |
|---|---|---|---|---|
| Hydraulic pressure at fracturation (MPa) | 0.97 | 3.25 | 1.05 | 1.66 |
| Estimated permeability (Darcy) | 0.001 to 0.004 | 0.04 | - | 0.1 |
| Calculated permeability (Darcy) | 0.028   −0.0028   −0.0023<br>−0.0028   0.024   −0.0012<br>−0.0022   −0.0011   0.034 | 0.046   −0.021   −0.014<br>−0.020   0.26   0.021<br>−0.012   0.020   0.041 | - | 0.333   0.055   0.0016<br>0.050   0.16   −0.046<br>0.0012   −0.061   0.068 |
| Angle between the fractures (°) | 178 | 165 | 167 | 134 - 88<br>(3 fractures) |

**Table 1.** *Four core samples were broken in the hydraulic cell. For all of them, a microtomography scan was acquired before and after fracturation.*

**3. Data analysis**

The permeability was estimated directly from the pressure drop and flux measured during the experiments. An alternative is to calculate the permeability tensor by solving the closure problem associated with the volume averaging of Stokes equations, as described by Whitaker (1996). Although computationally demanding, the solution of this closure problem can be calculated using established methods (Bernard *et al*. 2005).

Due to the large difference between the X-ray attenuation in the pores and in the grains, the grey-level distribution of the 3D images was almost bimodal. A simple threshold operation was applied to generate binary 3D images. This 3D image can be directly used as input geometry to perform permeability calculations. Only the pores forming a connected network between the two sides of the sample were kept for computing porosity.

This technique was employed on three subsamples (see Table 1), where the full permeability tensor could be calculated. For each sample, the trace of the tensor gives the permeability, which is shown to be of the same order of magnitude than the measured permeabilities. Using a Lattice Boltzmann technique to solve Stokes equations, Fredrich *et al.* (2006) have also shown that the permeability can be calculated within a factor 5 of error in a 3D porous domain. In our approach, the complete permeability tensor can be obtained.

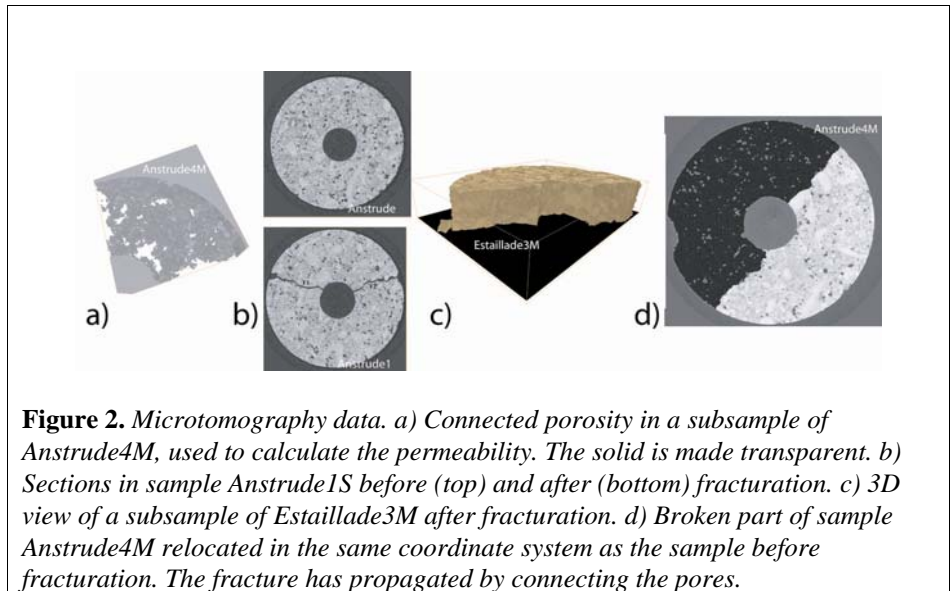

**Figure 2.** *Microtomography data. a) Connected porosity in a subsample of Anstrude4M, used to calculate the permeability. The solid is made transparent. b) Sections in sample Anstrude1S before (top) and after (bottom) fracturation. c) 3D view of a subsample of Estaillade3M after fracturation. d) Broken part of sample Anstrude4M relocated in the same coordinate system as the sample before fracturation. The fracture has propagated by connecting the pores.*

In order to highlight the fracture path in the rock, the 3D images of the two halves of the fractured sample were registered in the initial non fractured sample coordinate system (figure 2). This registration process first requires disconnecting the two halves of the broken sample (figure 2c) and processing them separately.

In the solid, some denser grains attenuate X-rays more than calcite grains, resulting in high intensity regions in the images. These areas can easily be segmented, and their centres of gravity are precisely located. As a result, a set of reference points was obtained for the unbroken sample, and for each half of the broken sample. Using these points and an iterative closest points algorithm, it was possible to calculate the translation and rotations that superimposes the point set (and thus the microtomography) of the initial and broken samples (Figure 2d).

Using this technique it was possible to localise with precision the path the fracture had followed. Preliminary results on sample Anstrude4M show that the fracture had connected the pores, and therefore broken the weak bonds between the grains. Note also that some grains have been broken.

## 4. Conclusions

A 3D simulation of permeability in porous limestones based on 3D real microstructure obtained by X-ray tomography has been performed. There is a reasonable agreement between the trace of the calculated permeability tensors and the experimentally measured permeability. The deviations can be attributed to the tendency for preferential flow channels in the sample compared to the subsamples used to calculate the permeability.

The microtomography data also show that the hydraulic fractures have propagated by connecting individual pores. Numerical simulations based on the deformation of a heterogeneous elastic solid under the effect of a pressure gradient should represent a future task for fracture propagation prediction.

*Acknowledgements*: The authors would like to thank Elodie Boller at ESRF, Robert Guiguet for designing the hydraulic cell, and Christophe Rousseau for his help with the hydraulic load control system and the data acquisition software.